\documentclass{aa}
\usepackage{graphics,epsfig,txfonts}

\usepackage{natbib}
\bibpunct{(}{)}{;}{a}{}{,}

\newcommand{\ergcm}[1]{$\times 10^{#1}$ erg cm$^{-2}$ s$^{-1}$}
\newcommand{\oergcm}[1]{$10^{#1}$ erg cm$^{-2}$ s$^{-1}$}
\newcommand{\ergs}[1]{$\times 10^{#1}$ erg s$^{-1}$}
\newcommand{\oergs}[1]{$10^{#1}$ erg s$^{-1}$}
\newcommand{\hcm}[1]{$\times 10^{#1}$ cm$^{-2}$}
\newcommand{\ohcm}[1]{$10^{#1}$ cm$^{-2}$}
\newcommand{\expo}[1]{$\times 10^{#1}$}
\newcommand{\oexpo}[1]{$10^{#1}$}

\newcommand{\lbol}{\hbox{L$_{\rm bol}$}}

\newcommand{\ct}{cts s$^{-1}$}

\newcommand{\Halp}{H${\alpha}$}
\newcommand{\ltsima}{$\buildrel < \over \sim$}
\newcommand{\lsim}{\lower.5ex\hbox{\ltsima}}
\newcommand{\gtsima}{$\buildrel > \over \sim$}
\newcommand{\gsim}{\lower.5ex\hbox{\gtsima}}
\newcommand{\xmm}{XMM-{\it Newton}}

\newcommand{\xmms}{\hbox{3XMM\,J051034.6$-$670353}} 
\newcommand{\xmmh}{\hbox{3XMM\,J051259.8$-$682640}} 
\def\rahour{\hbox{\ensuremath{^{\rm h}}}}
\def\ramin{\hbox{\ensuremath{^{\rm m}}}}
\usepackage{color}

\begin{document}
 
\title{EXTraS discovery of two pulsators in the direction of the LMC: \\
       a Be/X-ray binary pulsar in the LMC and a candidate double-degenerate polar in the foreground}

\author{     F. Haberl\inst{1}  
     \and    G. L. Israel\inst{2}
     \and    G. A. Rodriguez Castillo\inst{2}
     \and    G.~Vasilopoulos\inst{1}  
     \and    C.~Delvaux\inst{1}
     \and    A.~De~Luca\inst{3}
     \and    S.~Carpano\inst{1}
     \and    P.~Esposito\inst{4,3}
     \and    G.~Novara\inst{3,5}
     \and    R.~Salvaterra\inst{3}
     \and    A.~Tiengo\inst{3,5,6}
     \and    D.~D'Agostino\inst{7}
     \and    A.~Udalski\inst{8}  
       }

\titlerunning{EXTraS discovery of two pulsators in the direction of the LMC}
\authorrunning{Haberl et al.}

\institute{Max-Planck-Institut f\"ur extraterrestrische Physik, Giessenbachstra{\ss}e, 85748 Garching, Germany, \email{fwh@mpe.mpg.de}
            \and
           INAF-Osservatorio Astronomico di Roma, via Frascati 33, I-00040 Monteporzio Catone, Italy
            \and
           INAF–Istituto di Astrofisica Spaziale e Fisica Cosmica - Milano, via E. Bassini 15, I-20133 Milano, Italy
            \and
           Anton Pannekoek Institute for Astronomy, University of Amsterdam, Postbus 94249, NL-1090-GE Amsterdam, The Netherlands
            \and
           Scuola Universitaria Superiore IUSS Pavia, piazza della Vittoria 15, I-27100 Pavia, Italy
            \and
           INFN-Istituto Nazionale di Fisica Nucleare, Sezione di Pavia, via A. Bassi 6, I-27100 Pavia, Italy
            \and
           CNR--Istituto di Matematica Applicata e Tecnologie Informatiche, via de Marini 6, I-16149 Genova, Italy
            \and
           Warsaw University Observatory, Aleje Ujazdowskie 4, 00-478 Warsaw, Poland
          }

\date{Received 19 September 2016 / Accepted 10 October 2016}

 \abstract{The EXTraS project to explore the X-ray Transient and variable Sky searches for coherent signals in the X-ray archival data of \xmm.}
          {\xmm\ performed more than 400 pointed observations in the region of the Large Magellanic Cloud (LMC). We inspected the results of the EXTraS 
          period search to systematically look for new X-ray pulsators in our neighbour galaxy.} 
          {We analysed the \xmm\ observations of two sources from the 3XMM catalogue which show significant signals for coherent pulsations.}
          {3XMM\,J051259.8-682640 was detected as source with hard X-ray spectrum in two \xmm\ observations, revealing a periodic 
          modulation of the X-ray flux with 956~s. As optical counterpart we identify an early-type star with H${\alpha}$ emission. 
          The OGLE I-band light curve exhibits a regular pattern with three brightness dips which mark a period of $\sim$1350\,d. 
          The X-ray spectrum of 3XMM\,J051034.6-670353 is dominated by a super-soft blackbody-like emission component (kT $\sim$ 70 eV)
          which is modulated by nearly 100\% with a period of $\sim$1418 s. 
          From GROND observations we suggest a star with r' = 20.9 mag as possible counterpart of the X-ray source. 
          }
          {3XMM\,J051259.8-682640 is confirmed as a new Be/X-ray binary pulsar in the LMC.
          We discuss the long-term optical period as likely orbital period which would be the longest known from a high-mass X-ray binary.
          The spectral and temporal properties of the super-soft source 3XMM\,J051034.6-670353 are very similar to those of RX\,J0806.3+1527 
          and RX\,J1914.4+2456 suggesting that it belongs to the class of double-degenerate polars and is located in our Galaxy rather than in the LMC.}

 
\maketitle
 
\section{Introduction}
\label{sec:introduction}

One of the aims of the EXTraS (Exploring the X-ray Transient and variable Sky) project is to search systematically 
for coherent X-ray pulsations in the data obtained by the \xmm\ European Photon Imaging Cameras (EPIC), 
including pn \citep{2001A&A...365L..18S} and MOS type \citep{2001A&A...365L..27T} CCD detectors.
Selected from the \xmm\ EPIC Serendipitous Source Catalogue \citep[3XMM,][]{2016A&A...590A...1R}, the 
X-ray sources span six orders of magnitude in flux (10$^{-9}$ to \oergcm{-15} in the 0.2–12 keV energy band).
For further information about the EXTraS project, which also covers other aspects of characterising the 
temporal properties of the \xmm\ sources, see \citet{2016ASSP...42..291D} or the project web 
site\footnote{http://www.extras-fp7.eu}. The results from the EXTraS project including analysis software will 
be made public after the end of the project.

One of the first results from the EXTraS project was the discovery of a 1.2\,s modulation in the 
X-ray flux of a source located in an external spiral arm of our neighbour galaxy M\,31 \citep{2016MNRAS.457L...5E}. 
The source, 3XMM\,J004301.4+413017, is the first accreting neutron star found in M\,31 in which the spin period 
has been detected \citep[see also][]{2016arXiv160205191Z}. 
Many \xmm\ observations covered the source which revealed an orbital modulation 
at 1.27 d and made it possible to study the long-term properties of the source.

Inspired by the work on the X-ray pulsar in M\,31, we inspected the EXTraS results of the automated period search from 
all currently available archival \xmm\ observations (more than 400) in the direction of the Large Magellanic Cloud (LMC) 
to identify sources with hitherto unknown X-ray pulse periods.

\begin{table*}
\centering
\caption[]{\xmm\ observations of \xmmh\ and \xmms.}
\begin{tabular}{lcccrrrrr}
\hline\hline\noalign{\smallskip}
\multicolumn{1}{c}{Observation} &
\multicolumn{1}{c}{Start time} &
\multicolumn{1}{c}{Exp.} &
\multicolumn{1}{c}{Off-axis} &
\multicolumn{1}{c}{R.A. (DR6)} &
\multicolumn{1}{c}{Dec. (DR6)} &
\multicolumn{1}{c}{R.A. (SAS15)} &
\multicolumn{1}{c}{Dec. (SAS15)} &
\multicolumn{1}{c}{Err} \\
\multicolumn{1}{c}{ID} &
\multicolumn{1}{c}{} &
\multicolumn{1}{c}{} &
\multicolumn{1}{c}{angle} &
\multicolumn{2}{c}{(J2000)} &
\multicolumn{2}{c}{(J2000)} &
\multicolumn{1}{c}{} \\
\multicolumn{1}{c}{} &
\multicolumn{1}{c}{} &
\multicolumn{1}{c}{(s)} &
\multicolumn{1}{c}{(\arcmin)} &
\multicolumn{1}{c}{(h m s)} &
\multicolumn{1}{c}{(\degr\ \arcmin\ \arcsec)} &
\multicolumn{1}{c}{(h m s)} &
\multicolumn{1}{c}{(\degr\ \arcmin\ \arcsec)} &
\multicolumn{1}{c}{(\arcsec)} \\
\noalign{\smallskip}\hline\noalign{\smallskip}
\multicolumn{9}{c}{\xmmh} \\
  0690742301 & 2012-08-06 23:49 & 23407 & 16.7           & 05 13 00.03 & -68 26 38.6 & 05 13 00.30 & -68 26 39.2 & 1.4 \\
  0690742601 & 2012-08-12 23:27 & 24421 & \phantom{1}7.4 & 05 12 59.76 & -68 26 41.2 & 05 13 00.07 & -68 26 38.4 & 0.5 \\
\noalign{\smallskip}
\multicolumn{9}{c}{\xmms} \\
  0741800201 & 2014-05-17 02:36 & 31227 & 10.6           & 05 10 34.78 & -67 03 54.2 & 05 10 34.66 & -67 03 55.0 & 0.4 \\
  0741800301 & 2014-06-08 04:53 & 20407 & 10.8           & 05 10 34.51 & -67 03 52.5 & 05 10 34.62 & -67 03 54.1 & 0.5 \\
\noalign{\smallskip}\hline
\end{tabular}
\tablefoot{
Parameters are given for the EPIC pn instrument. Source coordinates are taken from the 3XMM-DR6 catalogue \citep{2016A&A...590A...1R}
and were also determined in this work using SAS\,15. 
A bore-sight correction using three QSOs was applied to the coordinates determined with SAS\,15 for observation 0741800301 (see Sect.~\ref{sec:soft_position}).
\xmmh\ was not covered by MOS1 during observation 0690742601 and neither by MOS1 nor MOS2 during observation 0690742301.
Similarly, \xmms\ was not covered by MOS1 in both observations. 
For our spectral and temporal analysis of \xmms\ we do not consider the MOS data due to the low efficiency in the soft band.
}
\label{tab:observations}
\end{table*}

A number of high-mass X-ray binary (HMXB) pulsars 
are known in this galaxy \citep[for a recent compilation see][]{2016MNRAS.459..528A} with several recent 
discoveries based on the \xmm\ survey of a $\sim$10 square degree area covering the central part of the LMC 
\citep{2013A&A...554A...1M,2013A&A...558A..74V,2014A&A...567A.129V,2016MNRAS.461.1875V}. As of today, pulse 
periods from 17 HMXBs in the LMC were published with periods between 0.069\,s and 2567\,s, with the bulk 
between 10\,s and 300\,s.
HMXBs are binary systems composed of a compact object and an early-type star. In a sub-group of HMXBs, the 
Be/X-ray binaries \citep[BeXRBs, e.g. ][]{2011Ap&SS.332....1R}, the compact object (typically a neutron star)
is accreting matter from the equatorial decretion disc surrounding a Be star \citep{2001PASJ...53..119O,2002MmSAI..73.1038Z}.
Pulsations detected in the X-ray flux from these systems indicate the spin period of the neutron star.

The LMC is also well known for its super-soft X-ray sources (SSSs), which form a quite heterogeneous class of objects 
\citep{1996LNP...472Q.299G,2008A&A...482..237K}. Many SSSs are luminous ($\sim$10$^{36}$ to $\sim$\oergs{38})
and have soft blackbody-like X-ray spectra with kT $\sim$ 20 $-$ 80 eV and can be detected in external galaxies. 
They can be explained by stable nuclear burning white dwarfs (WDs) accreting H-rich matter from a companion star
\citep{1992A&A...262...97V}. In M\,31, optical novae constitute the major class of SSSs 
\citep{2006A&A...454..773P,2014A&A...563A...2H}. 
Pulsations in the X-ray flux, suggesting the spin period of the WD, were detected in 
several SSSs in M\,31. The bright persistent SSS XMMU\,J004252.5+411540 shows a periodic modulation with a period of 
217\,s \citep{2008ApJ...676.1218T} in the X-ray emission which is most likely powered by steady burning of H-rich 
matter on the surface of the WD. Two other sources are likely associated with optical novae in their SSS state:
XMMU\,J004319.4+411759 with a period of 865\,s, which decreased significantly in flux (by a factor $\sim$8) 
between two \xmm\ observations about half a year apart \citep{2001A&A...378..800O}, and 
M31N 2007-12b, discovered in the optical during its nova outburst and detected by \xmm\ in its SSS state,
showing 1110\,s pulsations \citep{2011A&A...531A..22P}. The luminosity of M31N 2007-12b during maximum was 
around the Eddington limit of a massive WD and dropped by $\sim$30\% in an observation 60\,d after outburst.
The size of the photo-sphere with a radius of about 6000 km is consistent with emission from the full WD surface.

Soft blackbody-like emission is also observed from magnetic cataclysmic variables like
AM Her type systems \citep[polars,][]{1990SSRv...54..195C} or soft intermediate polars 
\citep{1995A&A...297L..37H,1996A&A...310L..25B,2008A&A...489.1243A}.
While in polars the rotation of the highly magnetised WD is synchronous with the orbital revolution (typical periods longer 
than 1 hour), the rotation period of WDs in intermediate polars can be as short as one thousands of the orbital period with 
typical spin periods of a few hundred seconds \citep{2006csxs.book..421K}. Double-degenerate binaries, i.e. systems hosting two 
interacting WDs, can have orbital periods as short as a few minutes. Such a framework is discussed for 
RX\,J1914.4+2456 \citep{1998MNRAS.293L..57C} and 
RX\,J0806.3+1527 \citep{2014A&A...561A.117E}.

In this paper, we present the discovery of X-ray pulsations from two sources located in the direction of the LMC, in the context  
of the EXTraS project. \xmmh\ is a known candidate for a BeXRB from positional coincidence but no pulse period was detected yet.
Apart from a possible detection with ROSAT, nothing was known about \xmms.
In Sect.~\ref{sec:x-ray} we describe the X-ray observations and the analysis methods
and in Sect.~\ref{sec:results} the results from our temporal and spectral analyses. We discuss our results 
in Sect.~\ref{sec:discussion}.

\section{\xmm\ observations and analysis}
\label{sec:x-ray}

From the inspection of the EXTraS results we found two sources from the 3XMM-DR6 catalogue \citep{2016A&A...590A...1R} located 
in the direction of the LMC which show significant (close to or above the 3.5$\sigma$ confidence threshold) pulsations. 
The first, \xmmh, was observed twice by \xmm\ in August 2012 within six days (see Table~\ref{tab:observations}), 
unfortunately covered only by part of the instruments because of the large off-axis angle in observation 0690742301 
and missing CCDs.
The second source, \xmms, was observed in May and June 2014. 

We re-processed the EPIC data using \xmm\ SAS 15.0.0\footnote{Science Analysis Software (SAS), http://www.cosmos.esa.int/web/ xmm-newton/sas}
to create event lists using the most recent software and calibration files. 
In particular, SAS\,15 includes a correction of the conversion between celestial and detector coordinate systems\footnote{http://www.cosmos.esa.int/web/xmm-newton/news-20160205}
while the 3XMM catalogue contains source positions determined with older SAS versions (up-to-date at the time of the observation).
We extracted the events to produce spectra and light curves from circular regions around the source positions and nearby blank-sky areas.
Single- and double-pixel events (PATTERN 0$-$4) were selected from pn data while we used singles to quadruples (PATTERN 0$-$12) from MOS, 
in both cases excluding bad CCD pixels and columns (FLAG 0). 
For the EPIC spectra we removed times of increased flaring activity when the background was above a threshold of 8 and 2.5 counts ks$^{-1}$ arcmin$^{-2}$ 
(7.0$-$15.0 keV band) for the EPIC-pn and EPIC-MOS detectors, respectively. To avoid time gaps in the time series we abstain from this technique
in the production of light curves. In particular the low energy bands are barely affected by the flares. 
EPIC spectra were re-binned to have at least 20 counts per bin and the SAS tasks {\tt arfgen} and  {\tt rmfgen} were used to generate 
the corresponding detector response files. The X-ray spectra were analysed with the spectral fitting package 
{\small XSPEC}\,12.9.0o\footnote{Available at http://heasarc.gsfc.nasa.gov/xanadu/xspec/} \citep{1996ASPC..101...17A}.
Errors are specified for 68\% confidence, unless otherwise stated.

\section{Results}
\label{sec:results}

\subsection{Temporal and spectral X-ray analysis}

The EXTraS pipeline to detect periodic signals in power spectra of time series from 3XMM sources is based on the algorithm described 
in \citet{1996ApJ...468..369I}. 
The pipeline uses archival event files selecting a standard energy band of 0.2 $-$ 10 keV. After the identification of new candidate pulsators, 
we performed a more detailed analysis of the sources using the re-processed event files.
To determine the most probable spin period together with the 1$\sigma$ uncertainty, we followed the Bayesian approach as described by 
\citet{1996ApJ...473.1059G} and applied by \citet{2008A&A...489..327H}.

\subsubsection{\xmmh}

\begin{figure}
\resizebox{\hsize}{!}{\includegraphics[angle=-90,clip=]{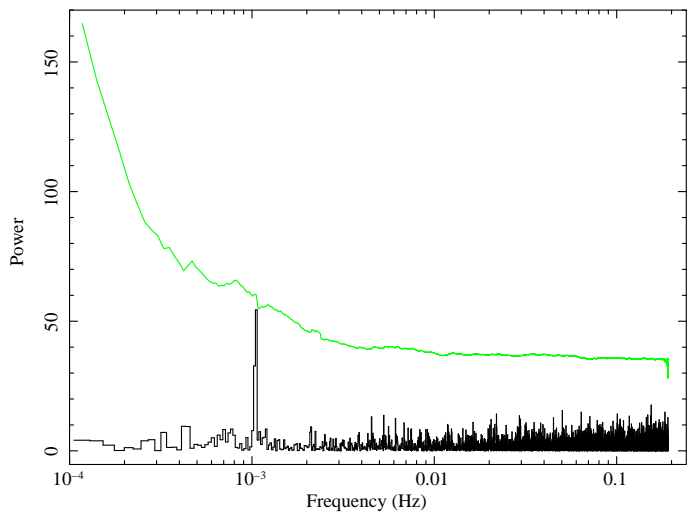}}
  
  \caption{
    Power spectrum for \xmmh\ obtained from EPIC pn data (observation 0690742601) produced by the EXTraS project in the 0.2$-$10 keV band.
    The green line indicates the 3.5$\sigma$ confidence level.
  }
  \label{fig:h_powspec}
\end{figure}

\begin{figure}
  \resizebox{\hsize}{!}{\includegraphics[angle=-90,clip=]{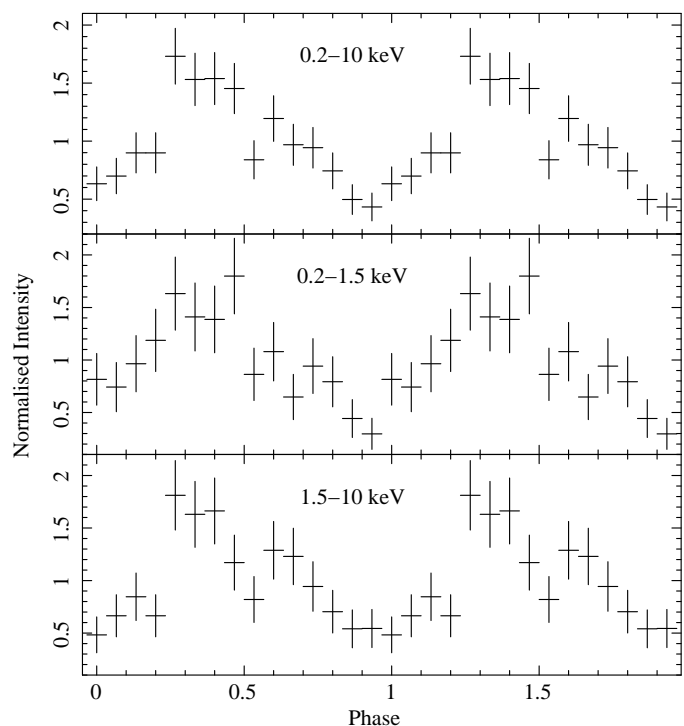}}
  \caption{
    EPIC pn light curves in different energy bands for \xmmh\ folded at a period of 956.2\,s (observation 0690742601).
  }
  \label{fig:h_foldlc}
\end{figure}

The power spectrum of \xmmh\ revealed a peak at a frequency near \oexpo{-3} Hz (Fig.~\ref{fig:h_powspec}). 
Using the EPIC-pn data in the 0.2$-$10 keV band from observation 0690742601 we determined the period to 956.2 $\pm$ 1.5 s. 
The folded light curve in the full energy band of 0.2$-$10 keV exhibits a slightly skewed sinusoidal profile 
(Fig.~\ref{fig:h_foldlc}) with a pulsed fraction of 47\% (uncertainty of 6\%, pulsed flux relative to the total).
Dividing the energies into two bands (0.2$-$1.5 keV and 1.5$-$10 keV) shows now significant differences in the pulse profiles
and pulsed fractions compatible within the uncertainties (60\% $\pm$ 8.5\% and 46\% $\pm$ 7.5\%, 
respectively). For a more detailed analysis of the energy dependence of the pulse profile, data with higher 
statistical quality are needed.

\begin{figure}
  \resizebox{\hsize}{!}{\includegraphics[angle=-90,clip=]{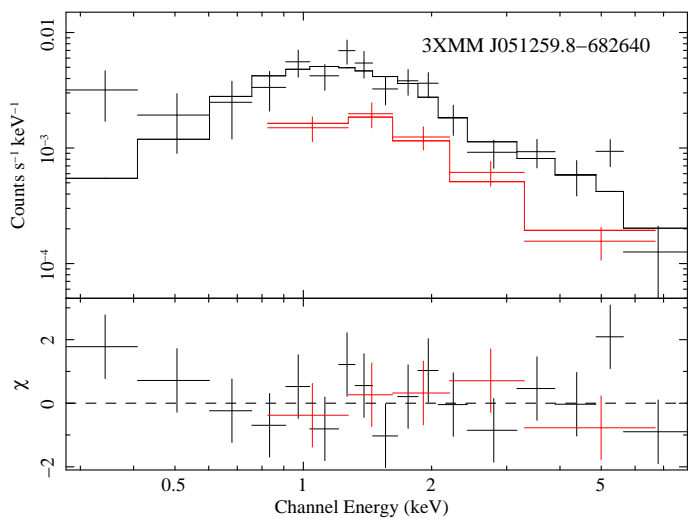}}
  \caption{
    EPIC pn and MOS2 spectra of \xmmh\ from observation 0690742601 together with the best fit model consisting of an absorbed power law.
  }
  \label{fig:h_pnspec}
\end{figure}

\begin{figure}
  \resizebox{\hsize}{!}{\includegraphics[angle=90,clip=]{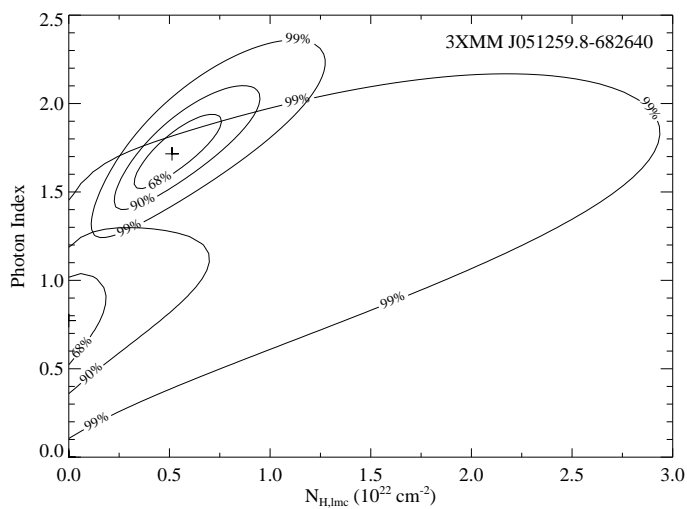}}
  \caption{
    Confidence contours for the LMC part of the column density, N$_{\rm H, lmc}$, and the photon index for \xmmh.
    The smaller contours are obtained from the second observation (0690742601).
  }
  \label{fig:h_pncont}
\end{figure}

We modelled the EPIC spectra of \xmmh\ with a simple power law, attenuated by two photoelectric absorption components
(model phabs*vphabs*powerlaw in XSPEC). For the foreground 
absorption in the Galaxy we assumed a column density of N$_{\rm H, gal}$ = 5.8\hcm{20} \citep[][]{1990ARAA...28..215D} 
with solar abundances according to \citet{2000ApJ...542..914W} and a free column density, N$_{\rm H, lmc}$, with abundances set to 
0.5 solar accounting for absorption by the interstellar medium of the LMC and local to the source. This model provides 
acceptable fits to the spectra of observation 0670742301 (pn) and 0670742601 
(simultaneous fit to pn and MOS2 with a normalisation factor - best fit value 1.06 - to allow for cross-calibration uncertainties, 
see Table~\ref{tab:spectral}). The spectra with best-fit model from observation 0670742601 are presented in Fig.~\ref{fig:h_pnspec}.
There is evidence for a spectral change between the two observations with a steeper spectrum during the later observation. Confidence 
contours for N$_{\rm H, lmc}$ and photon index for the two observations are compared in Fig.~\ref{fig:h_pncont}.

\subsubsection{\xmms}

\begin{figure}
  \resizebox{\hsize}{!}{\includegraphics[angle=-90,clip=]{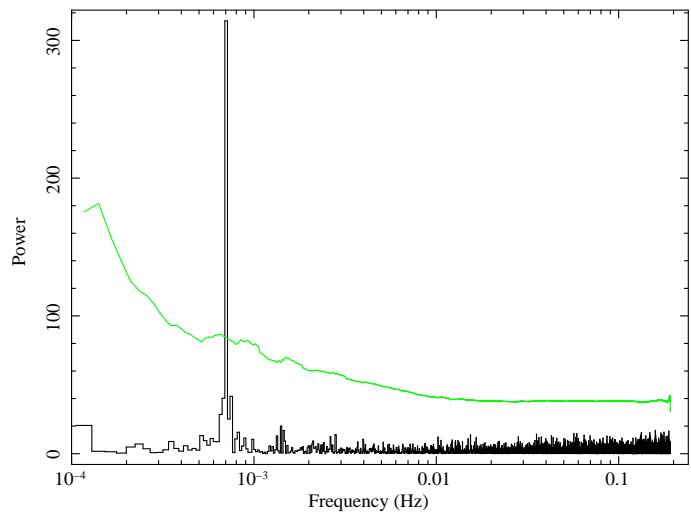}}
  \caption{
    As Fig.~\ref{fig:h_powspec}: Power spectrum for \xmms\ derived from observation 0741800201.
  }
  \label{fig:s_powspec}
\end{figure}

\begin{figure}
  \resizebox{\hsize}{!}{\includegraphics[angle=-90,clip=]{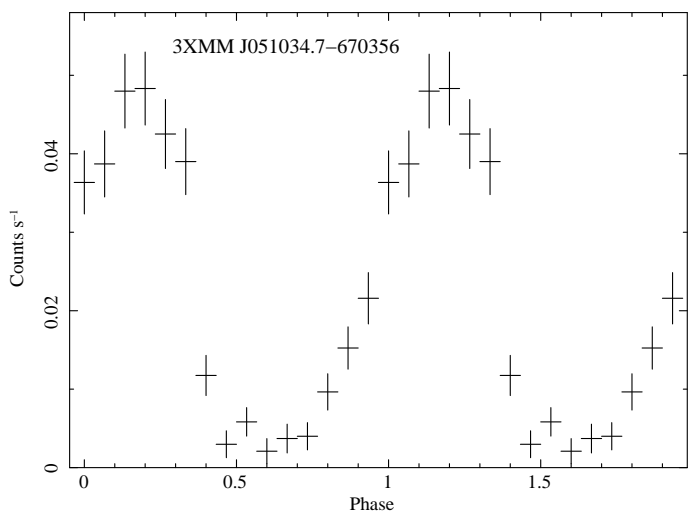}}
  \caption{
    Background-subtracted EPIC pn light curve of \xmms\ (observation 0741800201) in the 0.2$-$2 keV band folded at the period of 1417.4 s.
  }
  \label{fig:s_foldlc}
\end{figure}

The power spectra of \xmms\ revealed highly significant peaks at a frequency 
of $\sim$7.1\expo{-4} Hz (Fig.~\ref{fig:s_powspec}). Using the Bayesian technique, periods of 
1417.4 $\pm$ 1.1 s and 1419.4 $\pm$ 1.2 s were derived for observations 0741800201 and 0741800301, respectively.
The folded light curve (Fig.~\ref{fig:s_foldlc}) exhibits a deep modulation with little 
flux for more than one third of the period. 

\begin{figure}
  \resizebox{\hsize}{!}{\includegraphics[angle=-90,clip=]{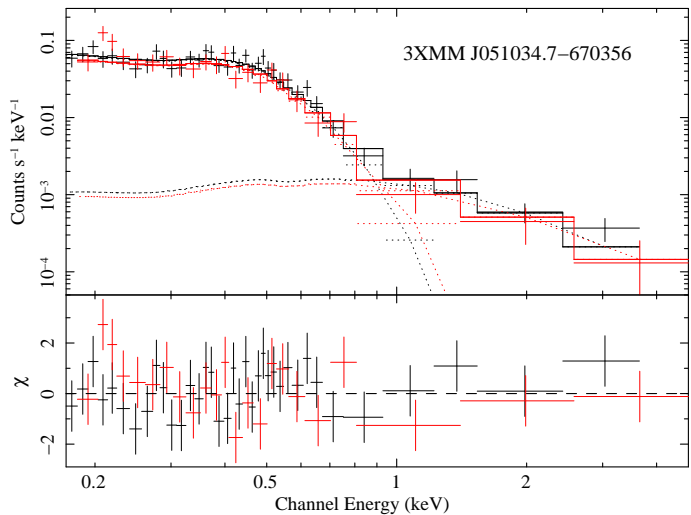}}
  \caption{
    EPIC pn spectra of \xmms\ (observation 0741800201 in black, 0741800301 in red) together with the best fit model (full lines) consisting of absorbed blackbody 
    and bremsstrahlung emission (dotted lines, dominating at low and high energies, respectively).
  }
  \label{fig:s_pnspec}
\end{figure}

\begin{figure}
  \resizebox{\hsize}{!}{\includegraphics[angle=90,clip=]{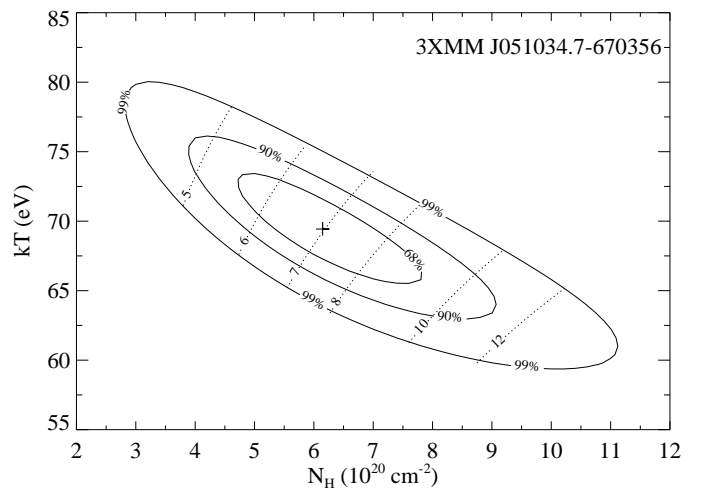}}
  \caption{
    Confidence contours for blackbody temperature and absorption column density derived from the EPIC pn spectrum of observation 0741800201,
    together with lines (dotted) of constant X-ray luminosity (0.1-2.4 keV, corrected for absorption).
    Luminosities are given in units of \oergs{32} for an assumed distance of 5\,kpc or \oergs{34} for a source located in the LMC at 50\,kpc.
  }
  \label{fig:s_pncont}
\end{figure}

We modelled the EPIC spectra of \xmms\ with blackbody emission
and a free column density with solar abundances. To account for a weak hard tail in the pn spectrum we added a 
bremsstrahlung component attenuated by the same amount of absorption (model phabs*(brems+bbody) in XSPEC). This
component is not well constrained and we fixed the temperature at 10 keV.
With this model, we fitted the spectra from the two observations individually and also simultaneously assuming no change in spectral shape, 
i.e. allowing only a constant factor between the spectra.
The EPIC pn spectra together with the simultaneous best-fit model are shown in Fig.~\ref{fig:s_pnspec}.
The best-fit parameters for both type of fits are listed in Table~\ref{tab:spectral}.
During observation 0741800301 the observed flux was lower by about 15\%, but with consistent spectral parameters.
This is consistent with the results of the simultaneous fit with acceptable quality of the fit for a constant factor of 0.84$\pm$0.10 (90\% confidence). 
Given the higher statistical quality due to higher flux and longer exposure time (31.2\,ks vs. 18.2\,ks after background flare screening)
we examine observation 0741800201 in the following.
In the 0.2 $-$ 10 keV band the observed flux of the bremsstrahlung component accounts for 29\% of the total, 
which decreases to 12\% when correcting for absorption, demonstrating the dominance of the soft component.
If we assume the source is located in the LMC, at a distance of 50 kpc, the 
bolometric luminosity of the blackbody component would 
be 1.05\ergs{35}, well below the range in which H-burning is expected to be stable \citep{2013ApJ...777..136W}. 
The blackbody luminosity depends somewhat on the temperature and absorption derived from the spectral fit, but should be at 
most a factor of $\sim$2 higher than the best-fit value (see Fig.~\ref{fig:s_pncont}).
In addition the radius of the emitting area of 180 km would be much smaller than the size of a WD.
On the other hand, the X-ray luminosity would be too high to be explained by a magnetic cataclysmic variable in the LMC.
If we assume the source to be located in the Milky Way, luminosity and radius of the emitting area are correspondingly
smaller (e.g. for distances of 1 to 5 kpc: \lbol\ = 4.2\expo{31} to 1.05\ergs{33} and r = 3.6 to 18 km).

\begin{table*}
\centering
\caption[]{Spectral fit results.}
\begin{tabular}{lcccccccc}
\hline\hline\noalign{\smallskip}
\multicolumn{1}{c}{Observation} &
\multicolumn{1}{c}{Photon} &
\multicolumn{1}{c}{kT$_{\rm bb}$} &
\multicolumn{1}{c}{N$_{\rm H}$} &
\multicolumn{1}{c}{$\chi^2_r$} &
\multicolumn{1}{c}{dof} &
\multicolumn{1}{c}{F$_{\rm observed}$\tablefootmark{a}} &
\multicolumn{1}{c}{L\tablefootmark{a,b}} &
\multicolumn{1}{c}{d\tablefootmark{c}} \\
\multicolumn{1}{c}{ID} &
\multicolumn{1}{c}{index} &
\multicolumn{1}{c}{} &
\multicolumn{1}{c}{} &
\multicolumn{1}{c}{} &
\multicolumn{1}{c}{} &
\multicolumn{1}{c}{} &
\multicolumn{1}{c}{} &
\multicolumn{1}{c}{} \\
\multicolumn{1}{c}{} &
\multicolumn{1}{c}{} &
\multicolumn{1}{c}{(eV)} &
\multicolumn{1}{c}{(\ohcm{21})} &
\multicolumn{1}{c}{} &
\multicolumn{1}{c}{} &
\multicolumn{1}{c}{(erg cm$^{-2}$ s$^{-1}$)}&
\multicolumn{1}{c}{(erg s$^{-1}$)} &
\multicolumn{1}{c}{(kpc)} \\
\noalign{\smallskip}\hline\noalign{\smallskip}
\multicolumn{9}{c}{\xmmh} \\
  0690742301 & 0.8 $\pm$ 0.5 & $-$         & 0.58\tablefootmark{d} / $<$7.0              & 0.74 & 10 & $1.4 \times 10^{-13}$ &  $4.1 \times 10^{34}$  & 50  \\
  0690742601 & 1.7 $\pm$ 0.4 & $-$         & 0.58\tablefootmark{d} / 5.1$^{+4.4}_{-2.9}$ & 0.92 & 18 & $7.5 \times 10^{-14}$ &  $3.2 \times 10^{34}$  & 50  \\
\multicolumn{9}{c}{\xmms} \\
  0741800201 & $-$           & 69 $\pm$  7 & 0.61$^{+0.29}_{-0.23}$                      & 0.80 & 33 & $9.2 \times 10^{-14}$ &  $7.5 \times 10^{32}$  & 5   \\
  0741800301 & $-$           & 78 $\pm$ 12 & $<$0.44                                     & 1.14 & 19 & $7.9 \times 10^{-14}$ &  $3.3 \times 10^{32}$  & 5   \\
  both       & $-$           & 72 $\pm$  6 & 0.45$^{+0.20}_{-0.16}$                      & 1.00 & 55 & $(9.2/7.7) \times 10^{-14}$ &  $(5.9/5.0) \times 10^{32}$  & 5   \\
\noalign{\smallskip}\hline
\end{tabular}
\tablefoot{Errors indicate 90\% confidence ranges.
\tablefoottext{a}{0.2 $-$ 10 keV.}
\tablefoottext{b}{Source intrinsic luminosity corrected for absorption.}
\tablefoottext{c}{Assuming \xmmh\ is located in the LMC \citep{2013Natur.495...76P} while the distance to \xmms\ is not known.}
\tablefoottext{d}{Foreground absorption in the Galaxy, fixed in the fit.}
}
\label{tab:spectral}
\end{table*}

\subsection{X-ray position and optical identification}

X-ray coordinates of our two sources provided by the 3XMM-DR6 catalogue are listed in Table~\ref{tab:observations}.
For comparison, we performed source detection using SAS\,15 following the standard procedures described 
in \citet{2013A&A...558A...3S}. The resulting positions differ by up to 3.7\arcsec, 
considerably larger than the statistical errors and indicating dominant systematic errors. 

\subsubsection{\xmmh}

The position of \xmmh\ in the 3XMM-DR6 catalogue (combining the two detections) is given as 
RA (J2000)=05\rahour12\ramin59\fs81 and Dec (J2000)=$-$68\degr26\arcmin40\farcs7 with an uncertainty of $\sim$0.7\arcsec.
This is $\sim$3\arcsec\ from likely counterparts found in optical catalogues.
Therefore, we considered only the EPIC data from observation 060742601 for our source detection with SAS\,15. 
During this observation the source was closer to the optical axes of the telescopes, providing a factor of two more net counts 
at better spatial resolution as compared to the other observation.
This X-ray position (Table~\ref{tab:observations}) is within 0.55\arcsec\ of 2MASS05125997$-$6826382 with J=14.23\,mag, H=14.00\,mag and K=13.72\,mag. 
This star is also listed in the MCPS catalogue of \citet{2004AJ....128.1606Z} with U=13.883\,mag, B=14.874\,mag and V=14.740\,mag,
resulting in the reddening-free Q parameter (Q = U$-$B$-$0.72$\times$(B$-$V)) of -1.087\,mag which together with optical brightness 
and colours is very typical for a BeXRB system \citep{2016A&A...586A..81H}. 
Moreover, the star is known as \Halp\ emission line star, [BE74] 226 \citep{1974A&AS...18...47B}, further supporting the 
identification of \xmmh\ as BeXRB.

\subsubsection{\xmms}
\label{sec:soft_position}

The soft X-ray source \xmms\ was detected in both \xmm\ observations, which covered the source.
The combined position in the 3XMM-DR6 catalogue is given with
RA (J2000)=05\rahour10\ramin34\fs64 and Dec (J2000)=$-$67\degr03\arcmin53.3\arcsec\ (error $\sim$0.74\arcsec).

A marginal ROSAT PSPC detection (existence likelihood 13.3) of a weak source with (5.7 $\pm$ 4.0)\expo{-3} \ct\ 
was found in the LMC catalogue of \citet[][entry 470]{1999A&AS..139..277H}.
The angular distance of 5.3\arcsec\ is well within the positional uncertainty of 38\arcsec\ of the PSPC detection.
The PSPC observation was performed between 1992-02-05 and 1992-04-28.
Using the spectral model parameters derived from the EPIC pn spectrum (see below) an expected PSPC count rate of 5.8\expo{-3} \ct\ was estimated.
Considering the relatively large error on the count rate of the PSPC detection this allows at most changes in flux by a 
factor of 3.5 (increase) or 1.7 (decrease) in 22 years, suggesting a relatively stable source flux on long time scales.

Within 3\arcsec\ of the 3XMM-DR6 position, only one star with V $\sim$ 21 is listed in the Magellanic Clouds Photometric Survey 
\citep[MCPS, ][angular separation 2.3\arcsec]{2004AJ....128.1606Z} and the OGLE catalogue \citep[][2.6\arcsec]{2000AcA....50..307U}. 
With B=21.11, V=20.81 and I=20.20 (MCPS) this object is relatively red.

\begin{figure}
  \resizebox{\hsize}{!}{\includegraphics[clip=]{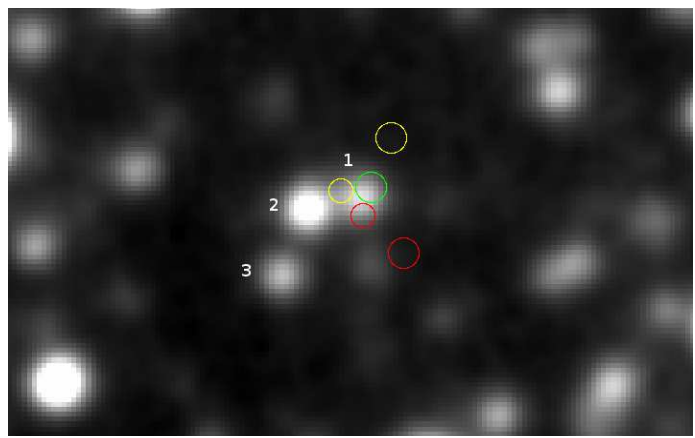}}
  \caption{
    GROND r' band image of the region around \xmms\ obtained on August 4, 2016 with an exposure time of 2160\,s.
    For each of the two \xmm\ observations the X-ray positions as provided in the 3XMM-DR6 catalogue (yellow circles) and 
    from our source detection with SAS\,15 (red circles) are marked.
    The green circle is centered on the position we derived (SAS\,15) from observation 0741800301 after astro-metric alignment using three QSOs.
    For the other observation no astro-metric correction was required.
    Positional errors (1$\sigma$) listed in Table\ref{tab:observations} are used for the circle radii. 
    For the three brightest objects closest to the X-ray positions (labelled Star 1, 2 and 3 counter clock wise with increasing distance)
    photometry was conducted in the seven GROND bands. 
  }
  \label{fig:grond_ima}
\end{figure}

To obtain deeper images of the area around \xmms, we performed observations with the Gamma-ray Burst Optical Near-ir Detector 
\citep[GROND,][]{2008PASP..120..405G} at the MPG 2.2\,m telescope in La Silla, Chile on 2016 August 4. 
2160\,s of integration time were obtained for the g' , r' , i' , and z' and 1800\,s for the J, H, and K bands. 
We analysed the data with the standard tools and methods described in \citet{2008ApJ...685..376K}.
Photometric calibration for the g' , r' , i' , and z' filter bands was obtained from an observation of an SDSS (Sloan Digital Sky Survey) standard star field. 
The J, H, and K photometry was calibrated using selected 2MASS stars \citep{2006AJ....131.1163S}. 
We derived the AB magnitudes (without correction for foreground reddening) for the three brightest objects near the X-ray 
position of \xmms\ (see Fig.~\ref{fig:grond_ima}) and list them in Table\ref{tab:grond_mag}.
In Fig.~\ref{fig:grond_ima} we also show for comparison the SDSS magnitudes \citep[data release 7,][]{2009ApJS..182..543A} 
for the double degenerate system RX\,J0806.3+1527 and for the soft intermediate polar PQ\,Gem. 

The position of the MCPS star (RA=05\rahour10\ramin34\fs94 and Dec=$-$67\degr03\arcmin54.7\arcsec) is consistent with ``Star 2''. 
The GROND colours are also consistent with a red object.
Among the three stars only ``Star 1'' shows a blue spectral energy distribution in the GROND filters which resembles 
that of RX\,J0806.3+1527, although not as steep as that of the likely double degenerate system.
The DR6 and SAS\,15 X-ray positions derived from observation 0741800201 are both compatible with  ``Star 1'' being the optical counterpart, 
while ``Star 3'' is probably too far away (Fig.~\ref{fig:grond_ima}). 
The coordinates derived from observation 0741800301 (DR6 and SAS\,15) are 2$-$3\arcsec\ away from ``Star 1'' and are not compatible with those 
from observation 0741800201. Therefore, we inspected the EPIC images and our SAS\,15 source detection lists to identify potential QSOs which can 
be used as reference system for astro-metric alignment. Using Vizier\footnote{http://vizier.u-strasbg.fr/viz-bin/VizieR}
we found three QSOs which all show a systematic shift between optical and X-ray positions derived from observation 0741800301 
(average +0.18\,s in R.A. and +2.1\arcsec\ in Dec.). This correction is applied to the coordinates given in Table~\ref{tab:observations}.
For observation 0741800201 the shifts are negligible with respect to the statistical errors and no correction was applied.
From positional coincidence, we consider ``Star 1'' as the most likely counterpart of the X-ray source.

\begin{table*}
\centering
\caption[]{GROND photometry of stars near \xmms.}
\begin{tabular}{lccccccc}
\hline\hline\noalign{\smallskip}
\multicolumn{1}{c}{Star} &
\multicolumn{1}{c}{g'} &
\multicolumn{1}{c}{r'} &
\multicolumn{1}{c}{i'} &
\multicolumn{1}{c}{z'} &
\multicolumn{1}{c}{J} &
\multicolumn{1}{c}{H} &
\multicolumn{1}{c}{K} \\
\multicolumn{1}{c}{} &
\multicolumn{1}{c}{(mag)} &
\multicolumn{1}{c}{(mag)} &
\multicolumn{1}{c}{(mag)} &
\multicolumn{1}{c}{(mag)} &
\multicolumn{1}{c}{(mag)} &
\multicolumn{1}{c}{(mag)} &
\multicolumn{1}{c}{(mag)} \\
\noalign{\smallskip}\hline\noalign{\smallskip}
 1 & 21.30 $\pm$ 0.01 & 21.33 $\pm$ 0.02 & 21.37 $\pm$ 0.04 & 21.48 $\pm$ 0.06 & $>$21.48         & $>$ 21.12        & $>$ 20.39 \\
 2 & 20.97 $\pm$ 0.03 & 20.88 $\pm$ 0.01 & 20.82 $\pm$ 0.02 & 20.75 $\pm$ 0.03 & 20.54 $\pm$ 0.13 & 20.39 $\pm$ 0.18 & $>$ 20.39 \\
 3 & 21.74 $\pm$ 0.02 & 21.61 $\pm$ 0.03 & 21.80 $\pm$ 0.06 & 21.72 $\pm$ 0.09 & $>$ 21.48        & $>$ 21.12        & $>$ 20.39 \\
\noalign{\smallskip}\hline
\end{tabular}
\label{tab:grond_mag}
\end{table*}

\begin{figure}
  \resizebox{\hsize}{!}{\includegraphics[angle=-90,clip=]{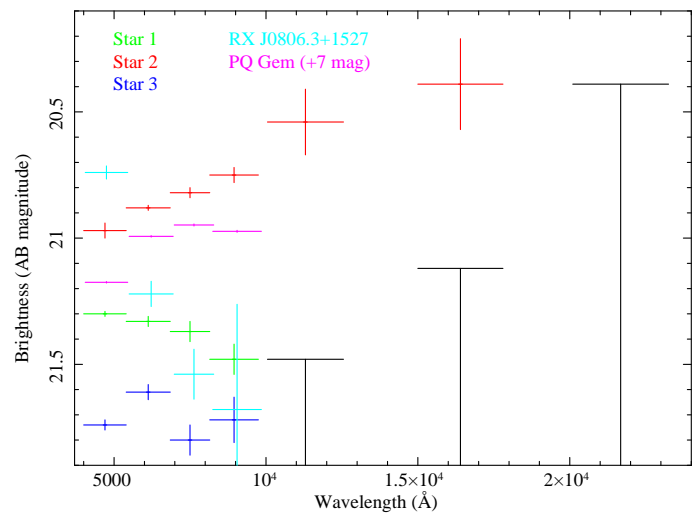}}
  \caption{
    GROND photometry for the three brightest stars (see Fig.~\ref{fig:grond_ima}) near the X-ray position of \xmms.
    The black symbols indicate upper limits (the same for all stars with non-detections).
    The SDSS magnitudes for RX\,J0806.3+1527 and PQ\,Gem (shifted by 7 magnitudes) are also plotted.
  }
  \label{fig:grond_mag}
\end{figure}

\subsection{OGLE data}

The regions around both of our X-ray sources were monitored regularly in the I- and V-band during phases III and IV of 
the Optical Gravitational Lensing Experiment \citep[OGLE;][]{2008AcA....58...69U,2015AcA....65....1U}. 
Images were taken in both V and I filter bands, while photometric magnitudes are calibrated to the standard VI system.
The optical counterpart of \xmmh\ and ``Star 2'' in the error circle of \xmms\ were covered by OGLE III and IV, while 
``Star 1'' unfortunately was located just outside of the CCD frame during phase IV. 

\subsubsection{\xmmh}

\begin{figure*}
  \resizebox{\hsize}{!}{\includegraphics[angle=-90,clip=]{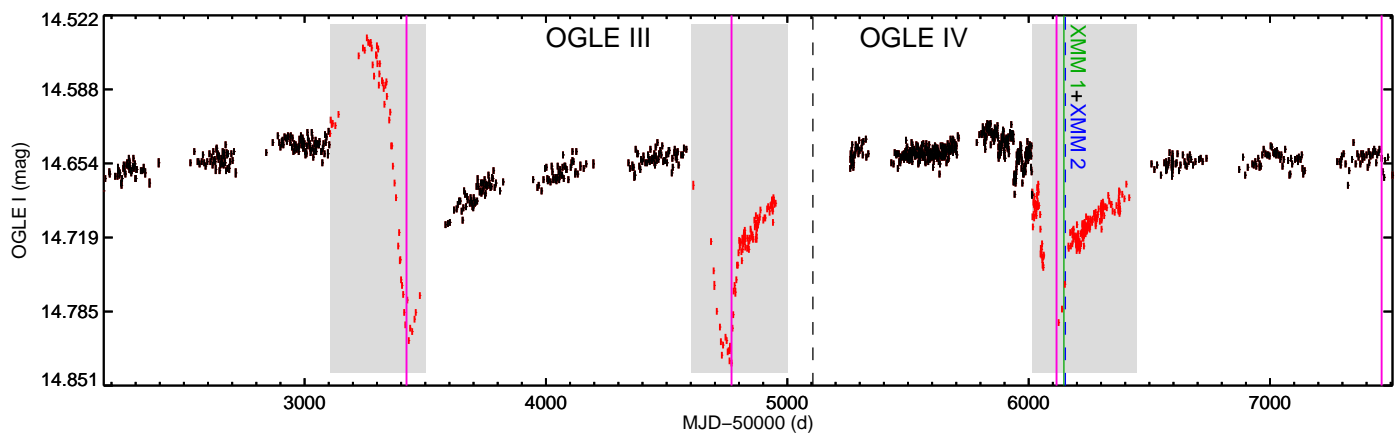}}
  \caption{
    OGLE I-band light curve of \xmmh\ between October 2001 and April 2016. 
    The vertical solid lines mark the minima of the dips with a recurrence time of 1350~d.
    The vertical dashed lines indicate the times of the \xmm\ observations.
  }
  \label{fig:h_oglelc}
\end{figure*}

From the optical counterpart of \xmmh\ almost 15 years of OGLE I-band data are available (see Fig.~\ref{fig:h_oglelc}).
The light curve exhibits a remarkable pattern with three deep intensity dips which are equally spaced by about 1350\,d 
and $\sim$100\,d wide.
The first of the dips was preceded by an increase in brightness by $\sim$0.1 mag in the I-band followed by a decrease 
by $\sim$0.25 mag. While the next two dips occurred without brightening, they reached about the same minimum of 14.8 mag,
in comparison to the level of $\sim$14.6 mag before the dips. After recovery from the last dip, the brightness reached 
only a level of 14.65 mag and no fourth dip is seen at the expected time. 

\begin{figure}
  \resizebox{\hsize}{!}{\includegraphics[clip=]{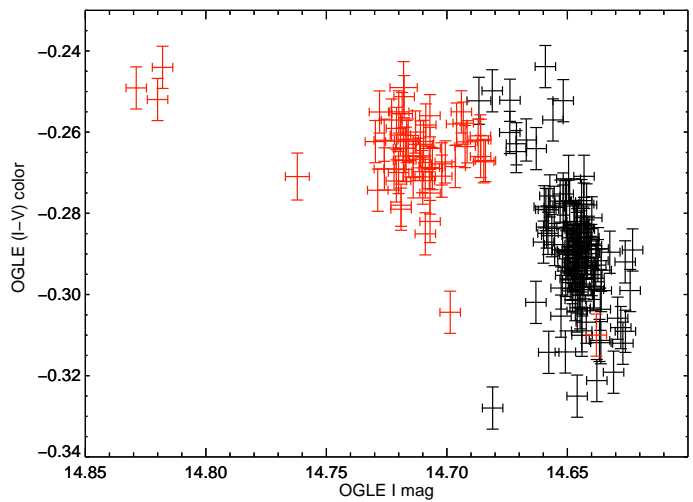}}
  \caption{
    OGLE colour (I$-$V) magnitude (I) diagram. Red crosses mark data points during the intensity dips as shown in Fig.~\ref{fig:h_oglelc}.
  }
  \label{fig:h_oglecol}
\end{figure}

OGLE V-band observations provided a less frequent coverage of the BeXRB system as compared to the I-band.
To compute the I$-$V colour index we directly used I- and V-band values when available from the same night.
Additionally, if I-band magnitudes were obtained less than four days before and after the V-band measurements, we 
interpolated the I-band magnitudes.
The colour index I$-$V is shown as function of the I magnitude in Fig.~\ref{fig:h_oglecol}. 
Data taken around the intensity dips are marked in red in Figs.~\ref{fig:h_oglelc} and \ref{fig:h_oglecol}.

\subsubsection{\xmms}

\begin{figure*}
  \resizebox{\hsize}{!}{\includegraphics[clip=]{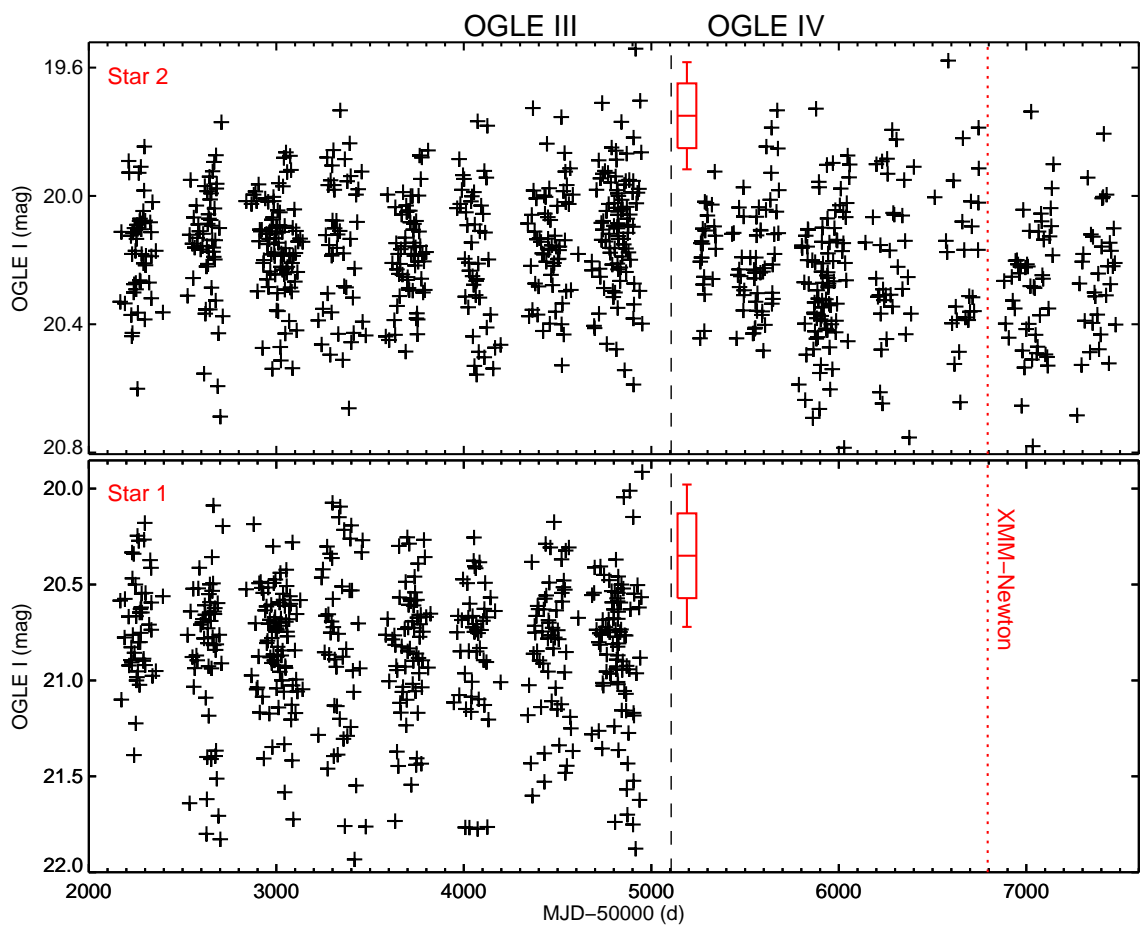}}
  \caption{
    OGLE light curves of the two candidate counterparts for \xmms\ since October 2001.
    The symbols in red indicate typical error bars with 50\% of the data points have errors smaller than the boxes, 
    while the emerging error bars enclose 75\% of the errors.
  }
  \label{fig:s_oglelc}
\end{figure*}

The OGLE I-band light curves of ``Star 1'' and ``Star 2'' are presented in Fig.~\ref{fig:s_oglelc}. 
The brightness of both stars is constant on time scales of years, but shows some scatter 
(2 mag and 1.2 mag for star 1 and 2 respectively) around the average that is larger than the 
typical error bar drawn in the figure.

\section{Discussion}
\label{sec:discussion}

From a systematic search for periodic variations in the X-ray flux of sources in the 3XMM catalogue within the 
EXTraS programme, we discovered two pulsators in the direction of the LMC.

\subsection{\xmmh}

The first pulsar, \xmmh, is characterised by a period of 956 s and a power-law X-ray spectrum. It can clearly be 
identified as new BeXRB pulsar in the LMC, making it the 18th high-mass X-ray binary pulsar in our neighbour galaxy. 
We identify the optical counterpart as an \Halp\ emission line star with brightness and colours typically found for BeXRBs in 
the Magellanic Clouds. The OGLE I-band light curve spans almost 15 years of data and shows three remarkable dips which are 
equally spaced by about 1350\,d (Fig.~\ref{fig:h_oglelc}). The regular occurrence of the three dips might indicate the 
orbital period of the binary system with the dips caused by interaction of the neutron star with the circum-stellar disk
when the neutron star approaches the Be star in an eccentric orbit. Between the dips, the optical brightness steadily 
increases, which in this picture can be explained by a recreation of the disk when the neutron star retreats.
The recreation might not always be complete as the missing dip (which should have occurred around mid March 2016) and 
the somewhat lower brightness level after the last dip suggest.
An orbital period of 1350\,d would be by far the longest known from an HMXB \citep[for the Galaxy see][]{2006A&A...455.1165L},
with several of the longest orbital periods reported for the Small Magellanic Cloud (SMC) of around 400 d \citep{2016A&A...586A..81H}. 
Only for CXOU\,J010744.51-722741.7 (= Swift\,J010745.0-722740) a similar period of 1180\,d, determined from recurring 
outbursts seen in the OGLE light curve, was suggested as orbital period \citep{2014ATel.5778....1M}.
Alternatively, the long period is more typical for a super-orbital period. These long-term periods, which are seen 
in most SMC BeXRBs, are likely caused by the formation and depletion of the circum-stellar disk \citep{2011MNRAS.413.1600R}.
However, the long-term variations seen in MACHO and OGLE data are more sinusoidal and do not exhibit the deep and sharp 
brightness dips seen from \xmmh.

During the brightness variations, the OGLE 
photometric data taken nearly simultaneously in the V and I bands show that the optical emission from the 
system becomes redder when increasing in brightness, similar to what is observed from other BeXRBs in the Magellanic 
Clouds \citep[e.g.][]{2014A&A...567A.129V,2012MNRAS.424..282C}.
The majority of BeXRBs with optical brightness variations in the SMC shows this behaviour  
with only four to five out of 31 becoming bluer with increasing brightness \citep{2011MNRAS.413.1600R}. 
To explain this, these authors suggest a geometrical viewing effect. The build-up of the circum-stellar disk 
leads to an overall reddening of the combined rising star/disk emission when the disk is viewed face-on, and a 
reddening when more of the (hotter) star is obscured (decreasing brightness) when the system is seen edge-on.
The colour evolution, bluest during the dips and redder with increasing brightness after the dips, is consistent
with disruption or truncation \citep{2001A&A...377..161O} and re-formation of the disk.
By chance the two \xmm\ observations of \xmmh\ were performed during the last dip seen in the I-band light curve.
The X-ray detection at a luminosity level of 4\ergs{34} during the optical minimum is at least consistent with 
the picture described above. Unfortunately, no sensitive observations were performed at other phases of the 
optical light curve. 
Upper limits from pointed ROSAT PSPC observations (assuming the spectral models derived from the EPIC pn spectra) 
are at a similar, or higher flux level as the \xmm\ detections.

\subsection{\xmms}

In contrast, \xmms\ shows a very soft X-ray spectrum which can be well modelled by absorbed blackbody 
emission and a weak hard emission component. 
The derived temperature of kT $\sim$ 70 eV is typical for soft emission seen from SSSs. In particular the LMC is 
well-known for its SSSs of various types \citep{2008A&A...482..237K}. However, the derived luminosity for an LMC distance 
is on one hand too low to be explained by steady H-burning on the surface of a WD \citep{2013ApJ...777..136W} 
and on the other hand too high to originate from a magnetic cataclysmic variable \citep[e.g.][]{2009A&A...496..121B}. 
Located closer within the Galaxy, the reduced X-ray luminosity would be well within the range of 
luminosities observed from cataclysmic variables and the small emission area is then 
consistent with that of a small fraction of the WD surface heated around the accreting poles. 

The pulse profile of \xmms\ is reminiscent to those of cataclysmic variables of type polar (AM Her) or 
intermediate polar \citep[e.g.][]{1995A&A...297L..37H}. 
While a pulse period of 1418 s is too short for a classical polar (with much longer synchronised WD-spin / orbital period),
it could be explained by the orbital period of an AM CVn system. 
In particular, the folded X-ray light curve of \xmms\ shows little or no flux for about one third of the pulse period, very similar
to RX\,J0806.3+1527 and RX\,J1914.4+2456 which are both discussed as double-degenerate polars.
\citet{2014A&A...561A.117E} estimate a distance of 0.9 kpc to RX\,J0806.3+1527 and report an unabsorbed X-ray flux of 1.5\ergcm{-12},
which is about a factor of 8 higher than we see from \xmms. Assuming that both sources have similar X-ray luminosity would place 
\xmms\ at a distance of $\sim$2.5 kpc, well within the old stellar population of the thick disk of the Milky Way \citep{2010IAUS..265..300B}.

A GROND observation of the field around \xmms\ revealed a possible counterpart (``Star 1'' in Fig.~\ref{fig:grond_ima}) which is the
closest of three r' = 20.9 to 21.6 mag objects to the X-ray positions derived from the two \xmm\ observations. The SED of ``Star 1''
in the optical GROND filter bands resembles that of RX\,J0806.3+1527, although less steep. 

Two main groups of models were proposed to explain the X-ray emission from double-degenerate polars. 
Accretion models which invoke mass transfer from a Roche-lobe filling WD predict a widening of the orbit, in 
contradiction to the observed decreases in orbital period from both RX\,J0806.3+1527 and RX\,J1914.4+2456
\citep[see][and references therein]{2014A&A...561A.117E}.
Alternatively, the ``unipolar inductor'' model, with a secondary WD that does not fill its Roche lobe and non-synchronous
rotation of the primary, has difficulties to provide enough energy by magnetic interactions to explain the observed X-ray luminosities 
\citep[equation 25 in][]{2012ApJ...757L...3L}. Given the strong dependence of the dissipation efficiency on the orbital period with 
L$_{\rm diss}$ $\propto$ P$^{-13/3}$, the maximum luminosity for \xmms\ is expected to be $\sim$1000 times smaller than that 
of RX\,J0806.3+1527. This would require that \xmms\ is much closer than RX\,J0806.3+1527 at $\sim$100 pc. This is inconsistent
with the faintness of the optical counterpart (I $\sim$ 20 mag or fainter). From the information we currently have, 
we conclude that \xmms\ is more likely powered by accretion. However, to further address this, we need to measure the orbital 
evolution of the binary system.
Whether \xmms\ indeed belongs to the class of double-degenerate systems with two interacting WDs, or to the soft intermediate polars, 
requires deep optical observations to search for a second period longer than 1418\,s. Orbital periods exceeding three hours are 
expected for soft intermediate polars (see Table~\ref{tab:softIP}) and if found, would identify the 1418\,s modulation as spin period of the 
white dwarf.

\begin{table*}
\centering
\caption[]{Cataclysmic variables with blackbody-like soft X-ray emission and fast spinning white dwarfs.}
\begin{tabular}{llcclc}
\hline\hline\noalign{\smallskip}
\multicolumn{1}{c}{Source} &
\multicolumn{1}{c}{Other name} &
\multicolumn{1}{c}{Periods} &
\multicolumn{1}{c}{kT [eV]} &
\multicolumn{1}{c}{Comment} &
\multicolumn{1}{c}{References} \\
\noalign{\smallskip}\hline\noalign{\smallskip}
 RX\,J0512.2-3241       & UU Col     & 863.5 s  / 3.45 h & 50        & soft IP             & 3,12,13   \\
 RX\,J0558.0+5353       & V 405 Aur  & 545.5 s  / 4.15 h & 57        & soft IP             & 2,4,13   \\
 1RXS\,J062518.2+733433 & MU Cam     & 1187 s   / 4.72 h & $\sim$43  & soft IP ?           & 8        \\
 1RXS\,J070407.9+262501 & V 418 Gem  & 480 s    / 4.37 h & 85        & soft IP             & 14,15    \\
 1E\,0830.9-2238        & WX Pyx     & 1557.3 s / $\sim$5.54 h & 82  & soft IP             & 10,13    \\
 RE\,0751+14            & PQ Gem     & 833.7 s  / 5.19 h & 52        & soft IP             & 1,2,5,13 \\
 1RXS\,J154814.5-452845 & NY Lup     & 693.0 s  / 9.87   & $\sim$100 & soft/hard IP        & 7,11,13   \\
 1RXS\,J180340.0+401214 & V 1323 Her & 1520.5 s / 4.4 h  & 100       & soft IP             & 14     \\
 RX\,J0806.3+1527       & HM Cnc     & 321.5 s           & 65        & double degenerate   & 9,16  \\
 RX\,J1914.4+2456       & V 407 Vul  & 567.7 s           & 43        & double degenerate   & 2,6   \\
 \xmms                  &            & 1418 s            & 69        & double degenerate ? & this work \\
\noalign{\smallskip}\hline
\end{tabular}
\tablefoot{References:
 1) \citet{1992MNRAS.258..749M}; 
 2) \citet{1995A&A...297L..37H}; 
 3) \citet{1996A&A...310L..25B}; 
 4) \citet{1996PASP..108..130S}; 
 5) \citet{1997MNRAS.288..817H}; 
 6) \citet{1998MNRAS.293L..57C}; 
 7) \citet{2002A&A...387..201H};
 8) \citet{2003A&A...406..253S};
 9) \citet{2003ApJ...598..492I};
10) \citet{2005A&A...433..635S};
11) \citet{2006A&A...449.1151D};
12) \citet{2006A&A...454..287D};
13) \citet{2007ApJ...663.1277E};
14) \citet{2008A&A...489.1243A}; 
15) \citet{2011PASP..123..130P}; 
16) \citet{2014A&A...561A.117E}
}
\label{tab:softIP}
\end{table*}

\begin{acknowledgements}
EXTraS is funded from the EU's Seventh Framework Programme under grant agreement no. 607452. 
This research is based on observations obtained with \xmm, an ESA science mission with 
instruments and contributions directly funded by ESA Member States and NASA. 
The OGLE project has received funding from the National Science Centre,
Poland, grant MAESTRO 2014/14/A/ST9/00121 to AU.
This research has made use of the VizieR catalogue access tool, CDS,
Strasbourg, France. The original description of the VizieR service was
published in A\&AS 143, 23.
GV acknowledges support from the BMWI/DLR grant FKZ 50 OR 1208 and
PE acknowledges funding in the framework of the NWO Vidi award A.2320.0076.
\end{acknowledgements}

\bibliographystyle{aa}
\bibliography{../../../../bibtex/general}

\end{document}